\documentclass[a4paper,11pt]{article}
\usepackage{pos}

\title{The Pierre Auger Observatory and Physics Beyond the Standard Model}
\ShortTitle{Pierre Auger Observatory and BSM physics}

\author*[a,b]{R. Aloisio}
\onbehalf{for the Pierre Auger Collaboration$^c$}

\affiliation[a]{Gran Sasso Science Institute,\\
viale F. Crispi n. 7, L'Aquila, Italy}

\affiliation[b]{INFN, Laboratori Nazionali Gran Sasso,\\
Via G. Acitelli, 22, Assergi L'Aquila, Italy}

\affiliation[c]{Observatorio Pierre Auger,\\
Av. San Mart\'in Norte 304, 5613 Malarg\"ue, Argentina.\\
Full author list \url{https://www.auger.org/archive/authors_2024_11.html}
}

\emailAdd{roberto.aloisio@gssi.it}

\abstract{The Pierre Auger Observatory, the world’s largest cosmic ray detector, plays a pivotal role in exploring the frontiers of physics beyond the standard model of particle physics. By the observation of ultra-high energy cosmic rays, Auger provides critical insights into two major scenarios: super heavy dark matter and Lorentz invariance violation. Super heavy dark matter, hypothesized to originate in the early universe, offers a compelling explanation for the dark matter problem and is constrained by Auger through searches for photons and neutrinos resulting from its decay. Lorentz invariance violations, motivated by quantum gravity theories implying deviations from fundamental symmetries, are probed by Auger through alterations of the particle dispersion relation and the energy thresholds of their interactions with astrophysical photons backgrounds.
}
\FullConference{7th International Symposium on Ultra High Energy Cosmic Rays (UHECR2024)\\
 17-21 November 2024\\
Malargüe, Mendoza, Argentina\\}

\begin{document}
\maketitle

\section{Introduction}

Ultra-high-energy cosmic rays (UHECRs), with energies exceeding $10^{20}$eV, are the most energetic particles ever observed. Their extreme energies provide a unique testing ground to search for new physics Beyond the Standard Model (BSM) of particle physics, that manifest at energies far beyond the current or foreseeable accelerator capabilities \cite{Adhikari:2022sve}. This includes Dark Matter (DM) models with superheavy particles, linking UHECR observations to the Dark Sector (DS) and the early-universe cosmology, and theories with Lorentz invariance violations, which connect UHECR observations to a broad class of quantum gravity models.
The Standard Model (SM) of particle physics is a theoretical framework remarkably successful at describing, with high accuracy, elementary particles and their interactions. However, it fails to address several key questions: such as the nature of dark matter and dark energy in the universe, the origin of neutrino masses and the naturalness problem (i.e. the fine-tuning needed in the SM to stabilize the electroweak scale against quantum corrections). In the last 40 years, these key questions have motivated extensive exploration of models BSM mainly in the framework of supersymmetric extensions, the reference approach to solve the naturalness problem which also provides natural candidates as DM particles: the so-called Weakly Interactive Massive Particles (WIMPs), with masses in the range of the electroweak scale $10^{2} \div 10^{4}$~GeV \cite{Bergstrom:2000pn}.

The possible connection between the solution of the naturalness problem and the DM problem triggered the strong hope that the search for WIMP DM could be connected with the discovery of super-symmetry at the TeV scale. Despite extensive experimental efforts through direct, indirect, and collider-based searches, no conclusive evidence for WIMPs or super-symmetry has been found \cite{Baer:2020kwz}. This fact calls for a paradigm change to address the DM problem. Recent analyses \cite{Buttazzo:2013uya}, based on the measurements at the Large Hadron Collider (LHC) of the masses of the Higgs boson and top quark, highlight that the running Higgs quartic coupling could approach zero at energy scales around $\Lambda_I=10^{10}\div 10^{12}$ GeV, signaling the possible emergence of a new ultra violet (UV) scale where an instability of the Higgs potential arises. This evidence, if confirmed, can be the first sign of new physics BSM at the LHC. The emergence of an UV scale $\Lambda_I$ in the SM might be associated to the DS of the theory, with the mass scale of DM particles corresponding to $\Lambda_I$. In this case the DM problem can be solved in the framework of Super Heavy Dark Matter (SHDM), particles with masses $M_X \gg 10^{3}$~GeV produced in the early universe.

On very general grounds, the viability of SHDM can be drawn under a few general hypotheses (see \cite{PierreAuger:2023vql,PierreAuger:2022ubv,PierreAuger:2022jyk,Guepin:2021ljb,Aloisio:2007bh} and refereces therein): (i) SHDM in the early Universe never reaches local thermal equilibrium; (ii) SHDM particles have mass much larger than the Higgs mass, linked to the specific DM production model in the early universe; and (iii) SHDM particles are long-lived particles with a lifetime exceeding the age of the Universe $\tau_{X} \gg t_0$. The abundance of SHDM today can be easily obtained in a wide class of SHDM production mechanisms, which naturally satisfy observational constraints on DM density without requiring fine-tuning (see \cite{PierreAuger:2023vql,PierreAuger:2022ubv,PierreAuger:2022jyk,Guepin:2021ljb,Aloisio:2007bh} and refereces therein). As for WIMPs, where the stability is assured by R-parity \cite{Bergstrom:2000pn}, in the case of SHDM  discrete gauge symmetries or weakly broken global symmetries can protect SHDM particles from decaying too rapidly, giving them lifetimes exceeding the age of the universe (see \cite{PierreAuger:2023vql,PierreAuger:2022ubv,PierreAuger:2022jyk,Guepin:2021ljb,Aloisio:2007bh} and refereces therein). These features make SHDM an attractive dark matter candidate. Observationally, SHDM decay produces UHECRs in the form of photons, neutrinos, and nucleons with a typical dominance of the neutral component \cite{Aloisio:2007bh}. Therefore, the key signature of SHDM decay is the observation of photons or neutrinos at the highest energies. The predicted energy spectrum of the secondary particles, sensibly harder than the underlying UHECR spectrum (see \cite{Aloisio:2007bh} and references therein), is another typical signature of SHDM decay. 

The extreme energies of UHECR are a test-bench for quantum gravity models, the highest energies observed as high as $10^{11}$ GeV are eleven orders of magnitude above the proton mass and "only" eight below the Planck mass $M_P=\sqrt{\tfrac{\hbar c}{G}}=10^{19}$ GeV/$c^2$. This is the mass scale where the Schwarzschild radius equals the Compton wavelength, and the gravitational and quantum regimes are expected to be at the same scale \cite{Addazi:2021xuf}. A quantum theory of General Relativity (GR) still eludes us, as it cannot be constructed as a renormalizable quantum field theory extension of GR, given the coupling constant of the theory, the Newton’s constant G, is not dimensionless (as the SM couplings). Practically all attempts to build a Quantum Gravity (QG) theory are based on the assumption that GR is an effective theory of a more fundamental (renormalizable) theory which arises at sufficiently low length scales around the Plank length $l_P=\sqrt{\tfrac{\hbar G} {c^3}} \simeq 10^{-35}$ m.  A general approach that can be followed to quantize gravity consists of considering space-time itself as a dynamical quantity. This approach dates back to the 1950s (see \cite{Addazi:2021xuf} and references therein) and shows that at sufficiently low length scales an uncertainty in the metric tensor itself arises, as quantum oscillations of the gravitational field make the geometry oscillate. These theories, such as loop quantum gravity or group field theories (see \cite{Addazi:2021xuf} and references therein), predict that spacetime could exhibit a discrete or foamy structure at the Planck scale, leading to small violations of Lorentz invariance. These effects are hypothesized to manifest through modifications to particle dispersion relations, resulting in energy-dependent variations in the speed of light or altered kinematics in high-energy processes. Such modifications may impact the thresholds for processes like photopion production and pair creation, directly affecting the propagation of UHECR, producing effects that can be detected by UHECR observatories, see \cite{PierreAuger:2021tog,Aloisio:2002ed} and references therein. 

The Pierre Auger observatory, located in Argentina, is the world’s largest UHECR detector, covering an area of 3,000 km$^2$. Its hybrid detection system combines surface water-Cherenkov detectors with fluorescence telescopes, enabling state of the art measurements of energy, mass, and arrival directions of UHECRs. In the present paper we will review the latest results of the Auger observatory \cite{PierreAuger:2023vql,PierreAuger:2022ubv,PierreAuger:2022jyk,PierreAuger:2021tog}, highlighting  the unparalleled sensitivity reached in testing SHDM and Lorentz invariance violation scenarios.

\section{Super Heavy Dark Matter models and constraints}

The basic parameters of any SHDM model, independent of the production and (quasi) stability mechanisms, are the particle's mass $M_{X}$ and lifetime $\tau_X \left (\gg t_0 \right )$. As discussed in the introduction, depending on the specific generation model, the mass scale of SHDM can vary over several orders of magnitude up to the Plank mass, as it can be in the range of $10^{8}~{\rm GeV}~ \leqslant M_X \leqslant 10^{19}$~GeV. The lifetime of SHDM should be discussed distinguishing between two alternative scenarios: with or without a direct coupling to the SM particles. In the case of a direct coupling, it should be extremely weak driven by a high energy scale $\Lambda>\Lambda_I$ as the Grand Unification Scale ($\Lambda_{GUT} \simeq 10^{16}$ GeV). 
\begin{figure}[!ht]
\centering
\includegraphics[scale=.28]{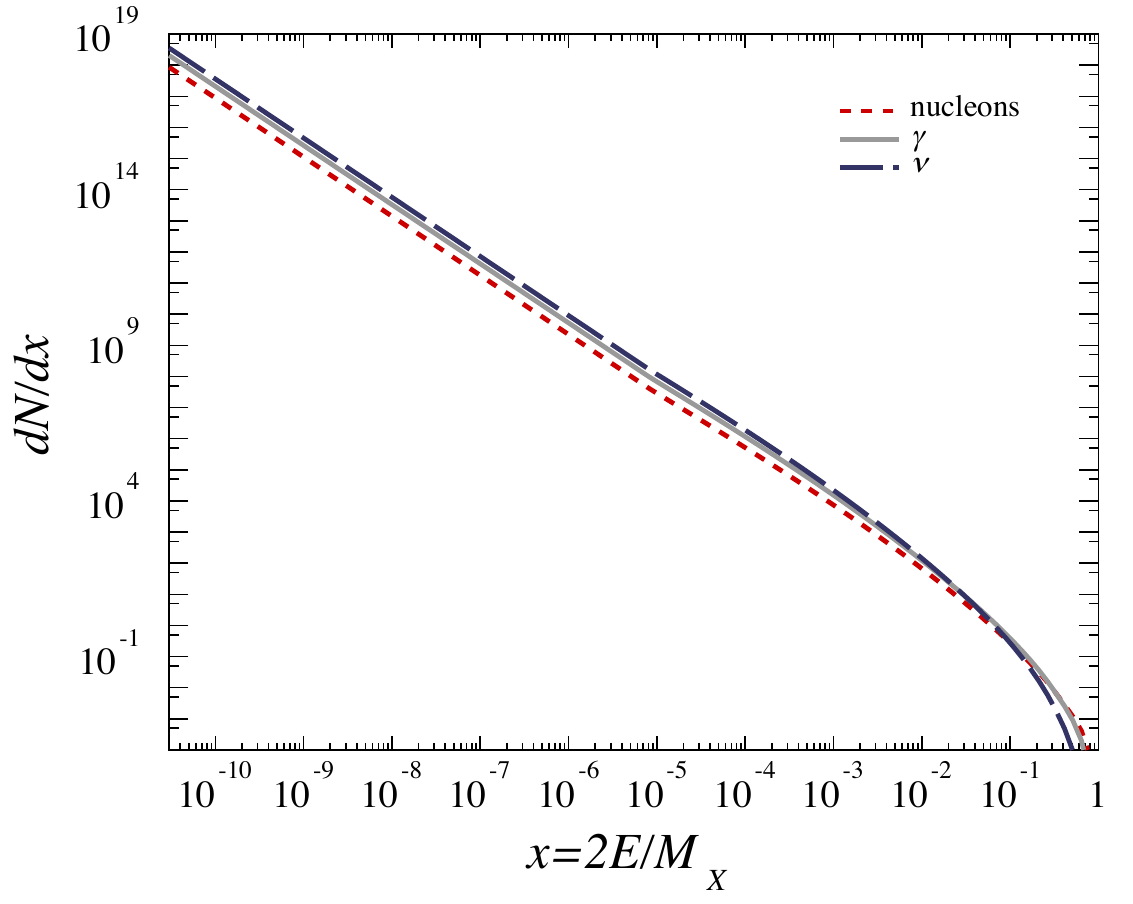}
\includegraphics[scale=.28]{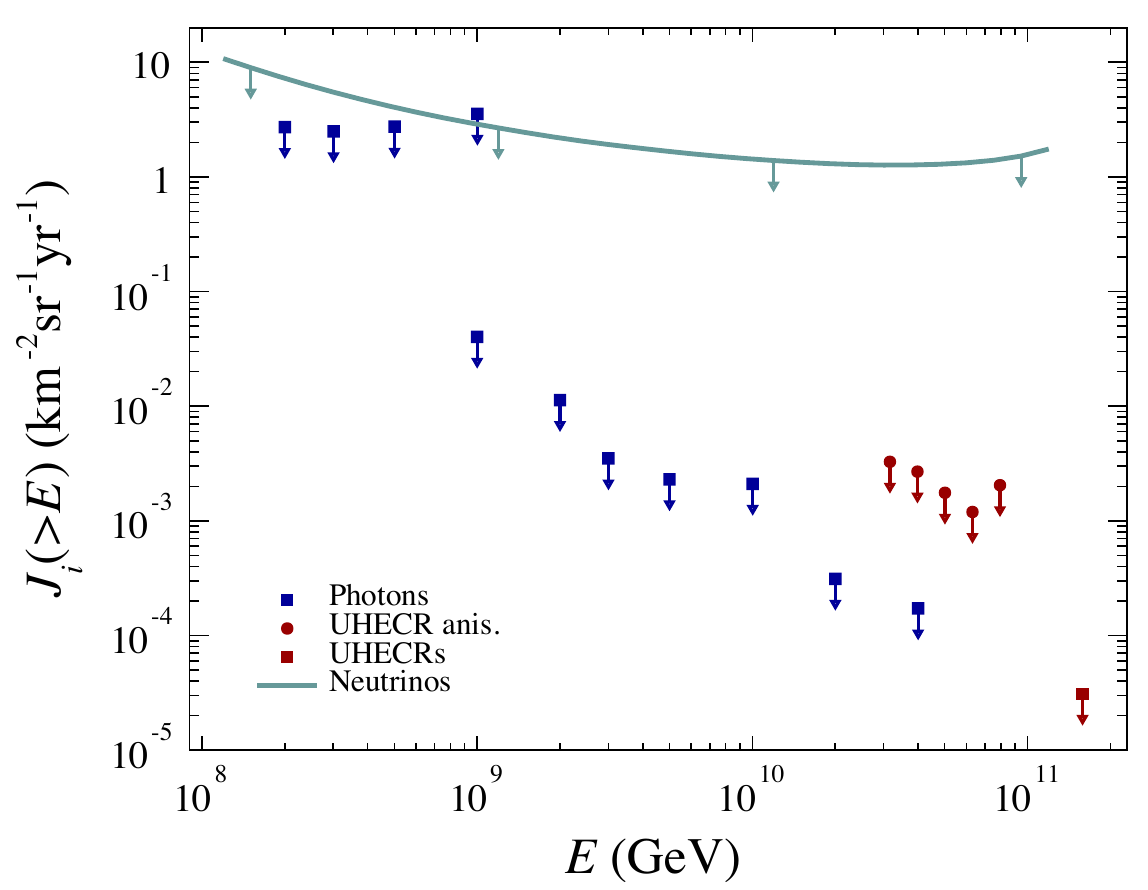}
\caption{ [Left Panel] Distribution function of the SM products in the SHDM decay. [Right Panel] Auger experimental limits on the fluxes of gamma rays and neutrinos. Figures taken from \cite{PierreAuger:2022ubv}}.
\label{fig1}
\end{figure}
\begin{figure}[!h]
\centering
\includegraphics[scale=.28]{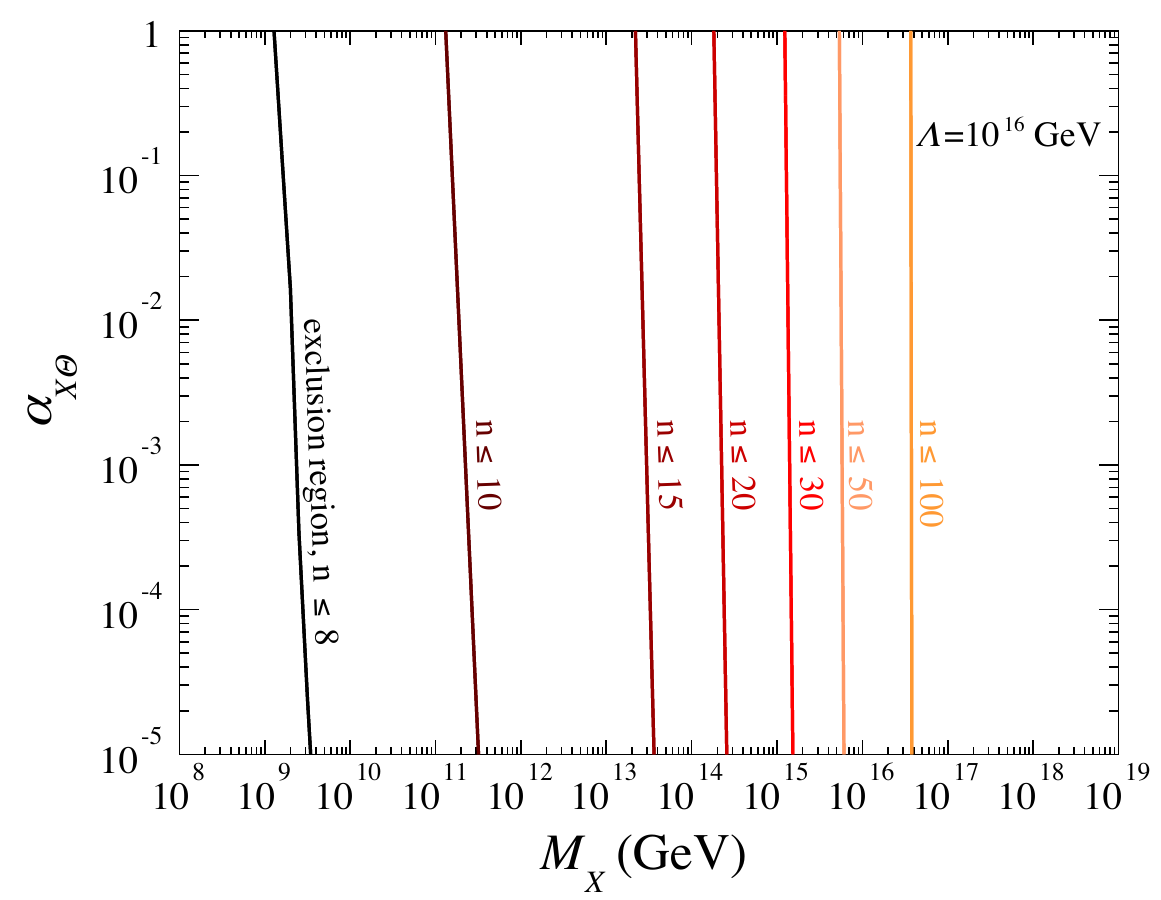}
\includegraphics[scale=.265]{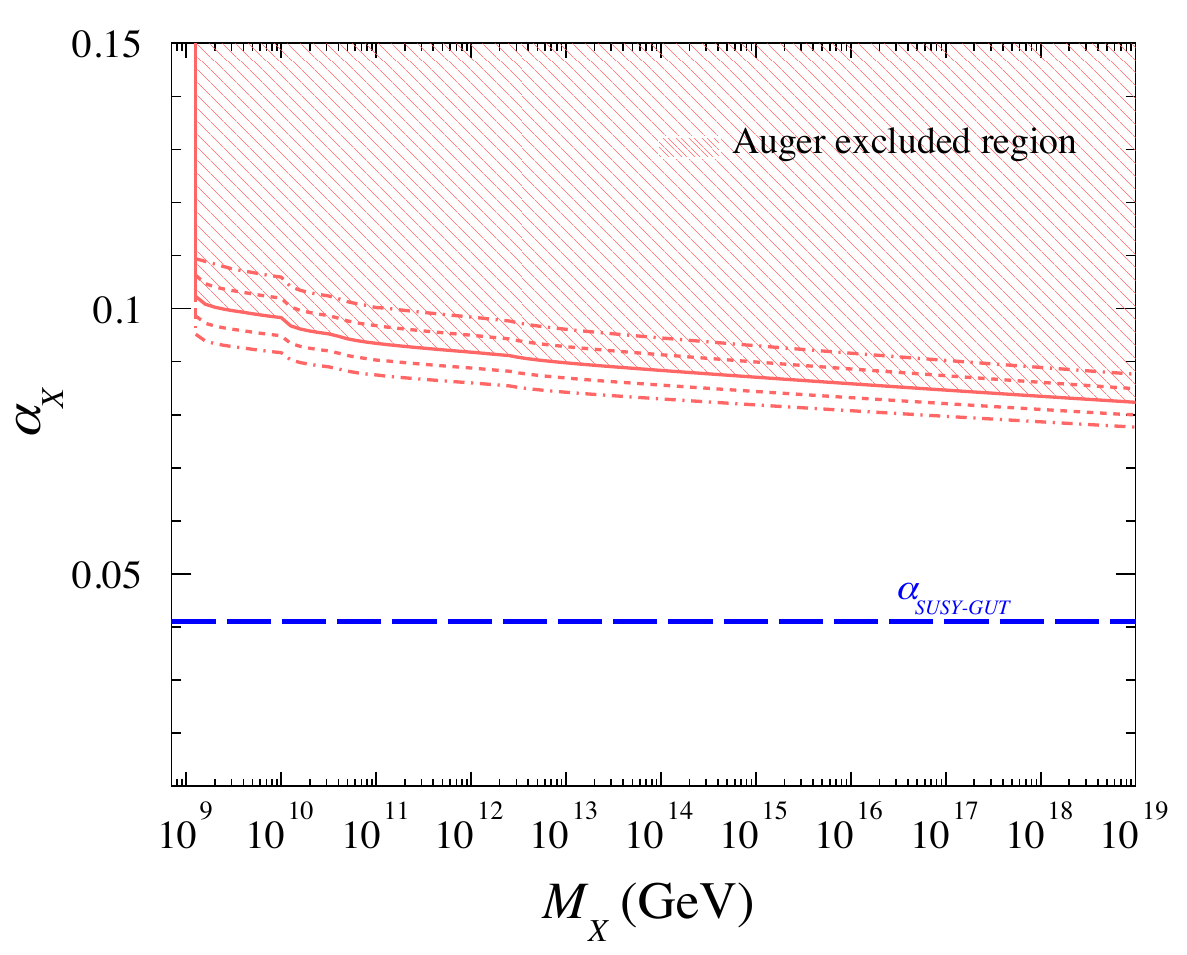}
\caption{Limits fixed by the Auger observations on the SHDM mass $M_X$ and: (left panel) the coupling SHDM-SM $\alpha_{X\Theta}$ for different values of the suppression scale order $n$, (right panel) the coupling of the hidden DS gauge interaction $\alpha_{X}$ in the case of the instanton-induced decay. Figures taken from \cite{PierreAuger:2022ubv}.}
\label{fig2}
\end{figure}
To assure long-lived particles, the interaction SHDM-SM should be suppressed by some power $n$ of the high energy scale $\Lambda$, with the lifetime $\tau_X\simeq (M_X \alpha_{X\Theta})^{-1} \left (\tfrac{\Lambda}{M_X} \right )^{2n-8}$, where $\alpha_{X\Theta}$ is the reduced coupling constant between SHDM and SM particles (see \cite{PierreAuger:2022ubv} and references therein). In the case in which SHDM interacts with SM particles only through the gravitational interaction\footnote{In this case a SHDM particle can be also called a Planckian-Interacting Massive Particle (PIDM) \cite{PierreAuger:2022ubv}.}, assuming the very general case of a DS characterized by its own non-abelian gauge symmetry, SHDM can decay only through non-perturbative effects like the instanton-induced decay (see \cite{PierreAuger:2022ubv} and references therein). In this case, the lifetime of SHDM follows from the corresponding instanton transition amplitude that, being exponentially suppressed, provides long-living particles. Considering the zeroth order contribution, the instanton-induced lifetime can be written as $\tau_X\simeq M_X^{-1} \exp{\left (\tfrac{4\pi}{\alpha_X} \right )}$, where $\alpha_X$ is the reduced coupling constant of the hidden gauge interaction in the DS. Under very general assumptions (see \cite{Adhikari:2022sve,PierreAuger:2022ubv,PierreAuger:2022jyk,Guepin:2021ljb,Aloisio:2007bh} and references therein), we can determine the composition and spectra of the standard model particles produced by the SHDM decay. 
\begin{figure}[!ht]
\centering
\includegraphics[scale=.28]{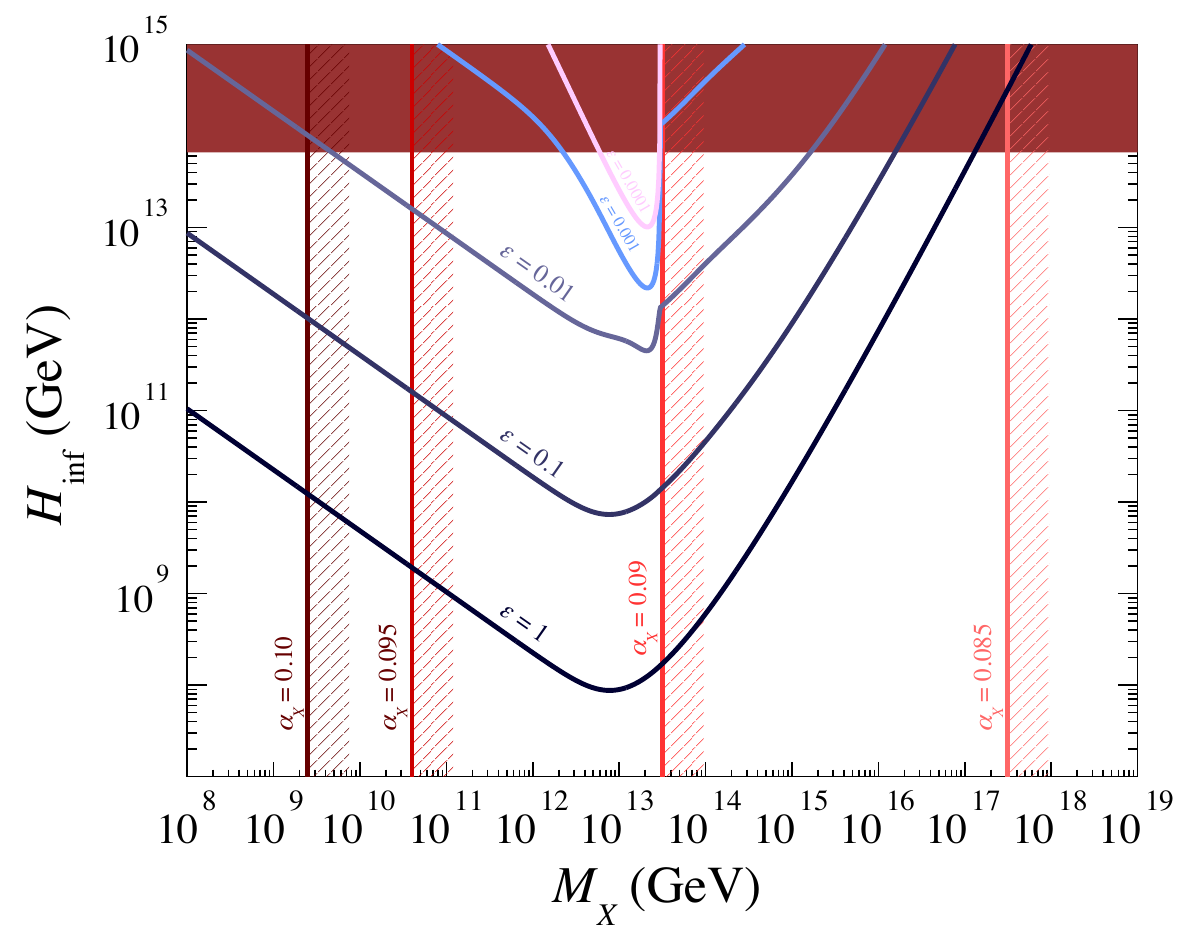}
\includegraphics[scale=.28]{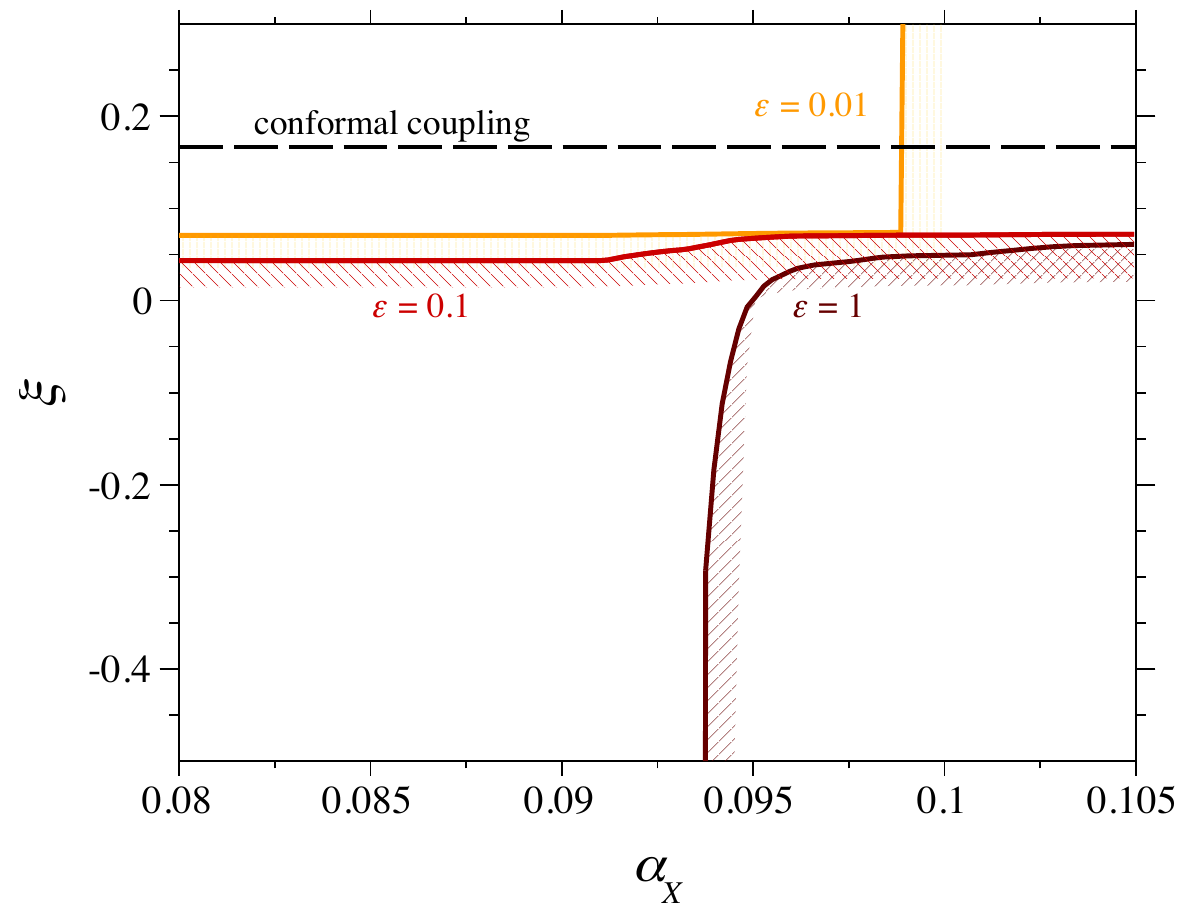}
\caption{[Left panel] The curves show the allowed parameters ($M_X,H_{inf}$) for different values of the reheating efficiency $\epsilon$ (see text). [Right panel] Stability regions of the Higgs potential in the ($\alpha_X,\xi$) plane, for different values of the reheating efficiency $\epsilon$ (see text).
Figures taken from \cite{PierreAuger:2022ubv}.}
\label{fig3}
\end{figure}
\begin{figure}[!ht]
\centering
\includegraphics[scale=.28]{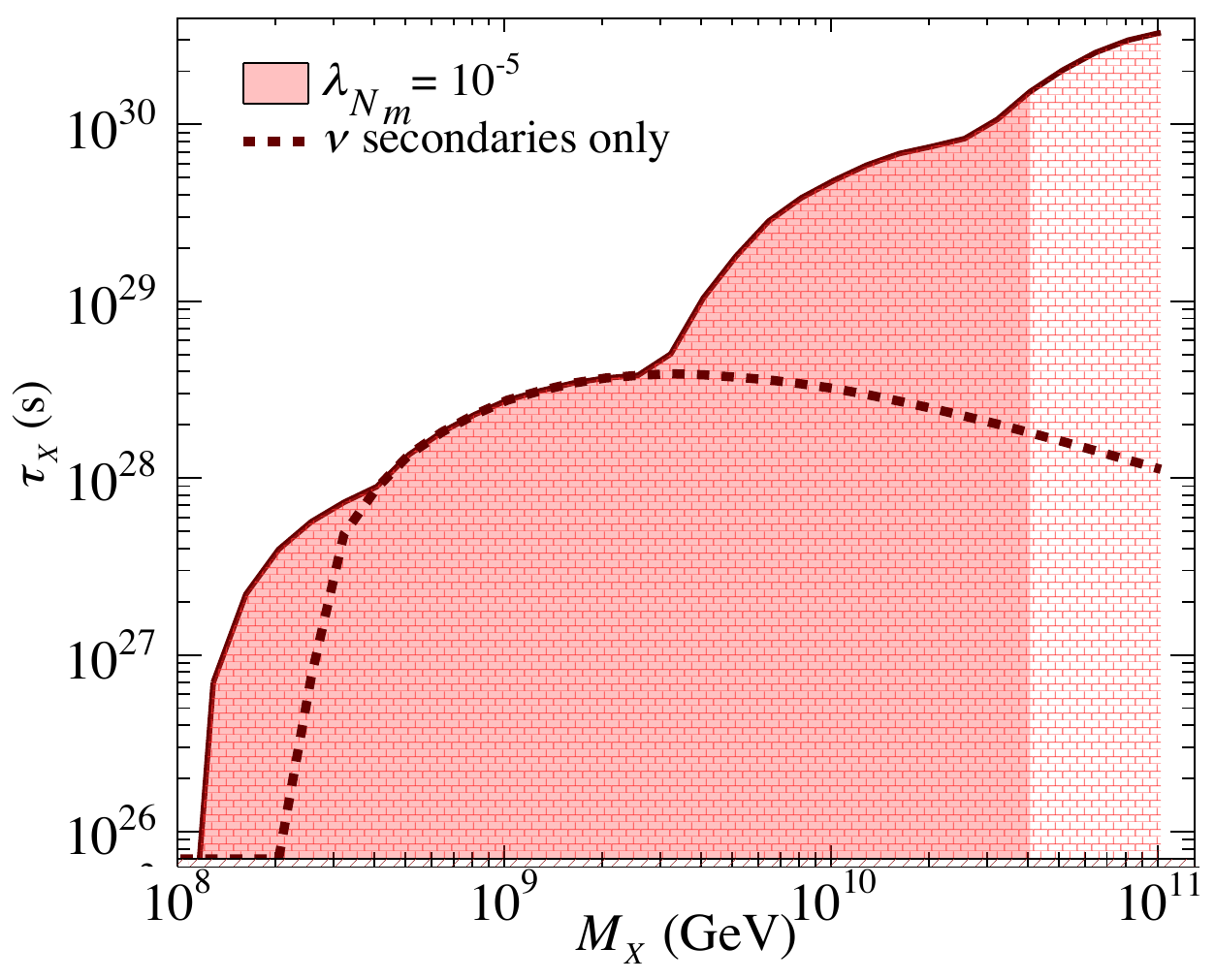}
\includegraphics[scale=.30]{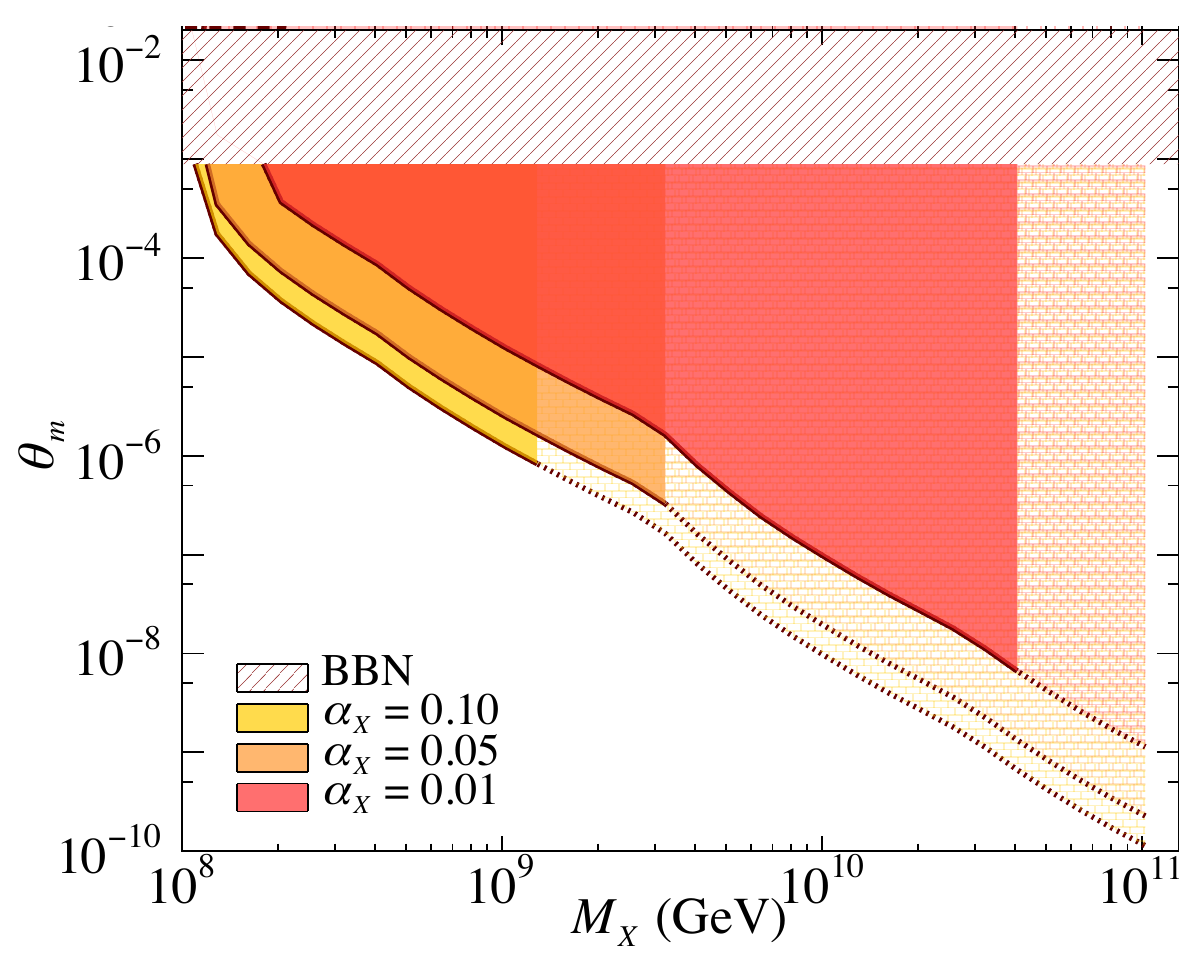}
\caption{[Left panel] Auger data excluded region ($M_X,\tau_X$) in the case of $X$ particles decaying into sterile neutrinos (model \cite{Dudas:2020sbq}) [Right panel] The same as in the left panel expressed in terms of the mixing angle $\theta_m$ between active and sterile neutrinos. Figures taken from \cite{PierreAuger:2023vql}.}
\label{fig4}
\end{figure}
Typical decay products are couples of quark and anti-quark\footnote{For a discussion of alternative (leptonic) decay patterns, see \cite{Guepin:2021ljb} and references therein.} that, through a cascading process (jets), give rise to neutrinos, gamma rays and nucleons. These particles exhibit a hard spectrum, that, as shown in the left panel of figure \ref{fig1}, can be approximated as $dN/dE \propto E^{-1.9}$, independently of the particle type, with a photon/nucleon ratio of about $\tfrac{\gamma}{N}\simeq 2\div 3$ and a neutrino nucleon ratio $\tfrac{\nu}{N}\simeq 3\div 4$, quite independent of the energy range \cite{Aloisio:2007bh}. Therefore, the most constraining limits on the SHDM mass and lifetime are those coming from the (non) observation of UHE photons and neutrinos in the UHE regime. As follows from the right panel of figure \ref{fig1}, Auger observations provide very stringent limits on the photon and neutrino fluxes in the energy range $10^{8}\div 10^{11}$ GeV, enabling by far the best limits on SHDM models \cite{PierreAuger:2022ubv,PierreAuger:2022jyk}. Assuming a DM density as in \cite{PierreAuger:2022ubv}, in figure \ref{fig2}, we plot the limits fixed by the Auger observations on the SHDM mass $M_X$ and the coupling SHDM-SM $\alpha_{X\Theta}$, in the case of direct coupling (left panel), and the gauge coupling $\alpha_{X}$ in the DS, for the instanton-induced decay (right panel). Requiring that the SHDM density today fits the observed DM density, in the framework of the models discussed in \cite{PierreAuger:2022ubv,PierreAuger:2022jyk}, the limits on $M_X$ and $\tau_X$ can be rewritten in terms of the cosmological parameters characterizing the generation mechanisms of SHDM, namely the age of the universe at inflation $H_{inf}^{-1}$ and the reheating efficiency $\epsilon=\left ( \tfrac{\Gamma_\phi}{H_{inf}} \right )^{1/2}$, where $\Gamma_\phi=\tfrac{g_\phi^2 M_\phi}{8\pi}$ is the inflaton (of mass $M_\phi$) decaying amplitude (see \cite{PierreAuger:2022ubv,PierreAuger:2022jyk} and references therein). In the left panel of figure \ref{fig3}, we plot on the plane ($M_X,H_{inf}$) the curves, for different values of $\epsilon$, along which the correct DM density today is recovered. 
The red area on the top signals the parameter region excluded by the limits on tensor to scalar modes in the Cosmic Microwave Background (CMB) \cite{Garny:2015sjg,Mambrini:2021zpp}. Additional constraints exclude the mass ranges in the regions to the right of the vertical lines, for the specified values of $\alpha_X$. In the right panel of figure \ref{fig3}, in the case of instanton induced decay, we plot the allowed regions of stability of the Higgs potential assuming a non-minimal coupling $\xi$ between the Higgs field and the space-time curvature during inflation (see \cite{PierreAuger:2022ubv,PierreAuger:2022jyk} and references therein). Given the limits on $\tau_X$, rewritten in terms of $\alpha_X$, the stability regions on the plane ($\alpha_X,\xi$) are those above the curves obtained for different values of $\epsilon$. The results presented in both panels of figure \ref{fig3} are very interesting, linking the observations of UHECR with those directly connected to cosmological models.  
A particular class of SHDM models, aiming at explaining neutrino masses and DM, with masses $M_X\geq 10^{9}$~GeV meet the lifetime requirements by coupling SHDM to a sector of sterile neutrinos (see the discussion in \cite{PierreAuger:2023vql}). In the reference model \cite{Dudas:2020sbq}, the SHDM is a pseudo-scalar particle $X$ with a Plank mass suppressed coupling to a sterile neutrinos sector. 
To produce the active neutrino masses the $X$ particle acts through a dynamical process similar to the Higgs mechanisms applied to right-handed neutrinos. A SHDM lifetime larger than the age of the universe is assured by the low ratio between the neutrino and Plank masses (see \cite{Dudas:2020sbq} and references therein). To assure the observed density of DM today in \cite{Dudas:2020sbq} it is assumed a coupling $\lambda_{N_m}\simeq 10^{-5}$ between the sterile neutrino sector and the inflationary one (see also \cite{PierreAuger:2023vql}). The dominant decay channel of the SHDM particle is a three body decay into the Higgs scalar and two sterile neutrinos $X\to h+\nu_1+\nu_2$, the relative decay width is $\Gamma_{h\nu_1\nu_2}^{X}=\tfrac{\alpha_X\theta_m}{192\pi^3}\left ( \tfrac{M_X}{M_P} \right )^2 \left (\tfrac{m_2}{m_{EW}} \right )^2 M_X$, where $\alpha_X$ is the coupling between the $X$ and sterile neutrino sectors, $\theta_m$ is the mixing angle between active and sterile neutrinos, $m_2$ is the mass of the sterile neutrino and $m_{EW}$ the electroweak scale. Neutrinos and photons are expected to be the final products of the decay. Using the Auger limits to the fluxes of these particles it is possible to severely constrain the model \cite{PierreAuger:2023vql}. In the left panel of figure \ref{fig4} we plot the excluded region in the plane ($M_X,\tau_X$) while in the right panel we plot the same limits in terms of the mixing angle $\theta_m$ \cite{PierreAuger:2023vql}.

\section{Lorentz Invariance Violations models and constraints}

As discussed in the introduction, different approaches to QG predict departures from Lorentz Invariance (LI) at extreme energy scales, typically at the Plank mass scale $M_P$ (see \cite{Addazi:2021xuf} and references therein). Effective field theories in which Lorentz Invariance is violated (LIV models) can be constructed assuming departures from the standard background metric at sufficiently high energy scales. 
\begin{figure}[!ht]
\centering
\includegraphics[scale=.59]{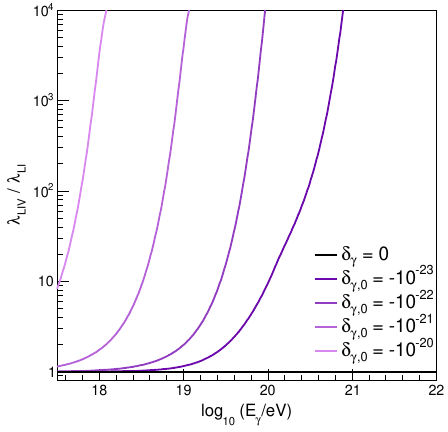}
\includegraphics[scale=.59]{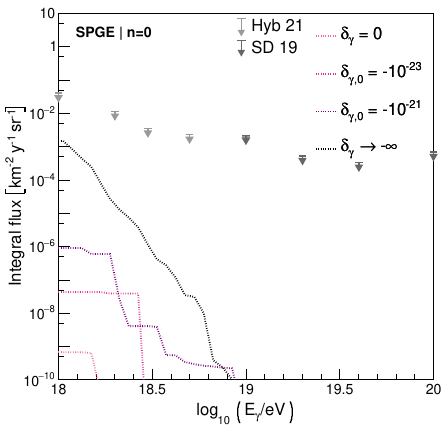}
\includegraphics[scale=.61]{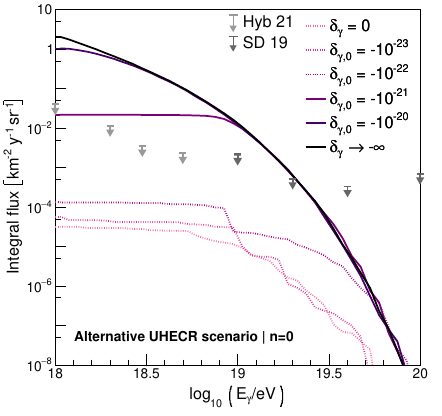}
\caption{[Left panel] Pair-production path length of UHE photons in the case of LI $\delta_\gamma = 0$ and for different values of the LIV parameter $\delta_{\gamma}$. [Central panel] Auger limits on gamma rays compared with gamma ray fluxes expected for different assumptions for the LIV parameter. [Right panel] As in the central panel with an additional subdominant proton component (see text). 
Figures taken from \cite{PierreAuger:2021tog}.}
\label{fig5}
\end{figure}
\begin{figure}[!ht]
\centering
\includegraphics[scale=.475]{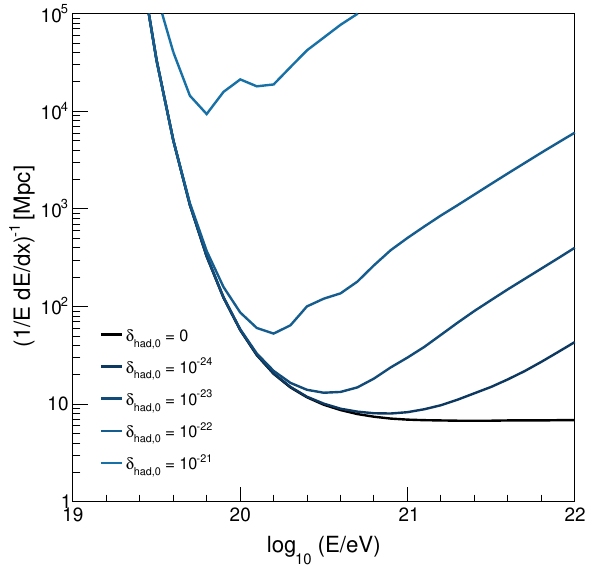}
\includegraphics[scale=.59]{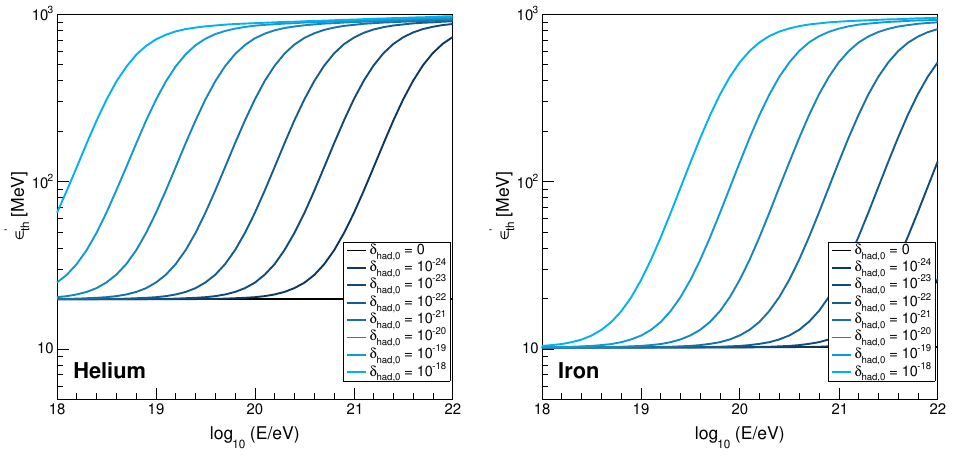}
\caption{[Left panel] Photo-pion production path-length of UHE protons in the case of LI $\delta_{had} = 0$ and for different values of the LIV parameter $\delta_{had}$. [Central panel] Photo-disintegration threshold (in the UHE nucleus reference frame) for helium nuclei in the case of LI $\delta_{had} = 0$ and for different values of the LIV parameter $\delta_{had}$. [Right panel] As in the central panel for iron nuclei. Figures taken from \cite{PierreAuger:2021tog}.}
\label{fig6}
\end{figure}
These departures, as discussed in \cite{Addazi:2021xuf,Coleman:1998ti}, imply modifications of the particle dispersion relation as $E^2-p^2=m^2\left [1 + g\left (\tfrac{E}{M_P}\right )\right ]+ E^2 f\left (\tfrac{E}{M_P}\right )$, with $f(0)=g(0)=0$. The LIV term being proportional to the particle mass (conformal LIV) is not testable experimentally and we will neglect it \cite{Addazi:2021xuf}. Expanding the modified dispersion relation around zero $(E \ll M_P)$ we can rewrite: $E^2-p^2=m^2+\sum_{n=0}^{\infty}\delta_nE^{2+n}$, where $\delta_n=\tfrac{\eta_n}{M_P^n}$ is the so-called LIV parameter \cite{PierreAuger:2021tog}.

As discussed in \cite{Addazi:2021xuf,PierreAuger:2021tog,Coleman:1998ti}, applying the LIV dispersion relations to the interaction processes that affect UHE particles propagation, by assuming momentum and energy conservations, only kinematical thresholds of the processes are affected. The relevant interaction processes are those between UHE particles and astrophysical photon backgrounds, such as the CMB and the Extragalatic Background Ligth, which give rise to the relevant processes: pair production by UHE photons $\gamma + \gamma_{bkg} \to e^+ + e^-$, photopion production by UHE protons $p+\gamma_{bkg}\to p+\pi^0(\pi^+\pi^-)$ and UHE nuclei photodisintegration $(A,Z)+\gamma_{bkg}\to (A',Z')+nN$. In figure \ref{fig5} we plot the UHE photon path-length in the case of standard LI propagation and assuming different values of the LIV parameter $\delta$. Relevant LIV scenarios are those with $\delta<0$ which imply an increase of the energy threshold for photon pair production. 
\begin{figure}[!ht]
\centering
\includegraphics[scale=.5]{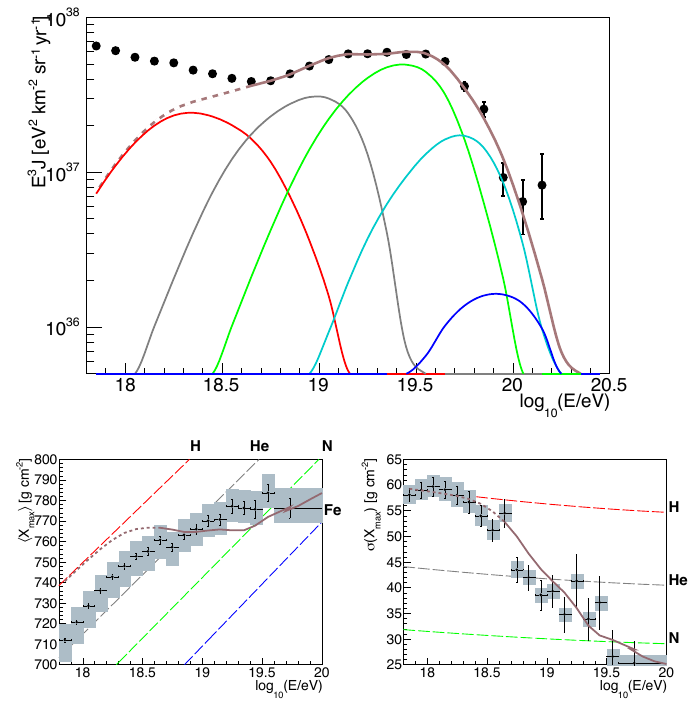}
\includegraphics[scale=.5]{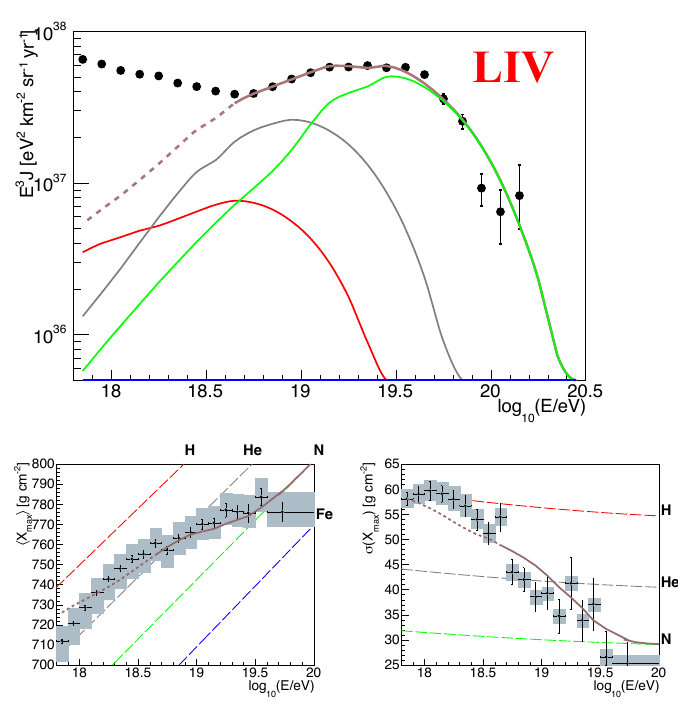}
\caption{[Left panel] Combined fit of the Auger spectrum and mass composition in the standard LI case \cite{PierreAuger:2016use}. [Right panel] As in the left panel for the LIV scenario \cite{PierreAuger:2021tog}.}
\label{fig7}
\end{figure}
In the central panel of figure \ref{fig5} we plot, for different values of $\delta$, the UHE photon flux produced by the process of photo-pion production by UHECR. The LIV scenario in this case is marginally constrained. To probe the maximum possible UHE photon flux expected in LIV scenarios, in the right panel of figure \ref{fig5} we plot the UHE photon flux obtained adding to the observed UHECR a (subdominant) proton component at the highest energies.    
In figure \ref{fig6} we consider the hadronic sector, which is affected by LIV parameters $\delta>0$. In the left panel we plot the proton path-length comparing the LI case with different assumptions for the LIV parameter. In the central and right panels, we plot the energy threshold of the photo-disintegration process (computed in the rest frame of the UHE particle). The effect of LIV can be therefore studied directly on the combined fit of the Auger data on spectrum and mass composition of UHECR \cite{PierreAuger:2016use}. In the right panel of figure \ref{fig7} we plot the combined fit of the Auger data in the standard LI case \cite{PierreAuger:2016use} and in the left panel the LIV case. By the comparison of LI and LIV case, using the Auger data, we can determine a very stringent limit on the LIV parameter at the level of $\delta < 10^{-20}$ \cite{PierreAuger:2016use}.


\begin{thebibliography}{99}


\bibitem{Adhikari:2022sve}
R.~X.~Adhikari,  \textit{et al.}
[arXiv:2209.11726].

\bibitem{Bergstrom:2000pn}
L.~Bergstr\"om,
Rept. Prog. Phys. \textbf{63}, 793 (2000)
[arXiv:hep-ph/0002126].

\bibitem{Baer:2020kwz}
H.~Baer,  \textit{et al.} 
Eur. Phys. J. ST \textbf{229} (2020) no.21, 3085-3141
[arXiv:2002.03013].


\bibitem{Buttazzo:2013uya}
D.~Buttazzo,  \textit{et al.} 
JHEP \textbf{12} (2013), 089
[arXiv:1307.3536 [hep-ph]].

\bibitem{PierreAuger:2023vql}
A.~Abdul-Halim \textit{et al.} [Pierre Auger],
Phys. Rev. D \textbf{109} (2024) L081101
[arXiv:2311.14541].

\bibitem{PierreAuger:2022ubv}
P.~Abreu \textit{et al.} [Pierre Auger],
Phys. Rev. D \textbf{107} (2023) no.4, 4
[arXiv:2208.02353].

\bibitem{PierreAuger:2022jyk}
P.~Abreu \textit{et al.} [Pierre Auger],
Phys. Rev. Lett. \textbf{130} (2023) no.6, 6
[arXiv:2203.08854].

\bibitem{Guepin:2021ljb}
C.~Gu\'epin, \textit{et al.}
Phys. Rev. D \textbf{104}, no.8, 083002 (2021)
[arXiv:2106.04446].


\bibitem{Aloisio:2007bh}
R.~Aloisio,  \textit{et al.}
Astropart. Phys. \textbf{29}, 307-316 (2008)
[arXiv:0706.3196].



\bibitem{Addazi:2021xuf}
A.~Addazi, \textit{et al.}
Prog. Part. Nucl. Phys. \textbf{125}, 103948 (2022)
[arXiv:2111.05659].


\bibitem{PierreAuger:2021tog}
P.~Abreu \textit{et al.} [Pierre Auger],
JCAP \textbf{01}, no.01, 023 (2022)
[arXiv:2112.06773].


\bibitem{Aloisio:2002ed}
R.~Aloisio, \textit{et al.} 
Astropart. Phys. \textbf{19}, 127-133 (2003)
[arXiv:astro-ph/0205271].

\bibitem{PierreAuger:2016use}
A.~Aab \textit{et al.} [Pierre Auger],
JCAP \textbf{04} (2017), 038
[arXiv:1612.07155].














\bibitem{Garny:2015sjg}
M.~Garny, \textit{et al.}  
Phys. Rev. Lett. \textbf{116} (2016) no.10, 101302
[arXiv:1511.03278].

\bibitem{Mambrini:2021zpp}
Y.~Mambrini and K.~A.~Olive,
Phys. Rev. D \textbf{103} (2021) no.11, 115009
[arXiv:2102.06214].



\bibitem{Dudas:2020sbq}
E.~Dudas,  \textit{et al.} 
Phys. Rev. D \textbf{101} (2020) no.11, 115029
[arXiv:2003.02846].

\bibitem{Coleman:1998ti}
S.~R.~Coleman and S.~L.~Glashow,
Phys. Rev. D \textbf{59} (1999), 116008
[arXiv:hep-ph/9812418].









\end{thebibliography}
\end{document}